# On the two-dimensional Brillouin flow

**Ryan Revolinsky, Christopher Swenson, Nicholas Jordan, Y.Y. Lau[a], and Ronald Gilgenbach**

Plasma, Pulsed Power, and Microwave Laboratory

Department of Nuclear Engineering and Radiological Sciences, University of Michigan, Ann Arbor, MI 48109

[a]Corresponding author: yylau@umich.edu

**ABSTRACT**

The Brillouin flow is a rectilinear, sheared electron fluid flow in a crossed electric field (E) and magnetic field (B), in the E × B direction with zero flow velocity and zero electric field at the surface with which the flow is in contact. It is broadly considered as the equilibrium electron flow in high power crossed-field devices including the magnetron and magnetically insulated transmission line oscillator. This paper provides an examination of Brillouin flow in two dimensions, in a cylindrical geometry where the anode radius changes abruptly at a single axial location while the cathode surface has a constant radius. Our simulation confirms the proof that there is no equilibrium Brillouin flow solution for such a geometry. It further reveals that this change in the anode radius introduces novel bunching of the electrons within the Brillouin hub. This bunching occurs at low frequencies and is very pronounced if the Brillouin flow is from the small gap region to the large gap region, but is minimal if the Brillouin flow is from the large gap region to the small gap region. New insights are provided into the physical processes that initiate and sustain the bunching processes that are unique for a crossed-field diode, as compared with a non-magnetized diode. We argue that this enhanced bunching, and its concomitant formation of strong vortices, is not restricted to an abrupt change in the anode-cathode gap spacing.

**I. INTRODUCTION**

Brillouin flow is a one-dimensional (1D) fluid flow in a magnetically insulated crossed-field diode. It is a laminar, rectilinear shear flow of electrons within a crossed-field gap with zero flow velocity and zero electric field at the surface with which the flow is in contact. Virtually all equilibrium and stability studies of the Brillouin flow assume an equilibrium in one-dimension (1D), whether the magnetic insulation of the equilibrium flow is provided by an external magnet, as in magnetron and relativistic magnetron, or by the wall currents without an external magnet, as in magnetic insulated line oscillator (MILO) and magnetically insulated transmission line (MITL). The electrons generated from the cathode surface of these devices, whether by thermionic emission, field emission, or other means, initially execute cycloidal orbits that eventually evolve into a Brillouin-like flow if there is a steady supply of electrons from the cathode surface. The Brillouin flow has long been conjectured to be the preferred state over the cycloidal orbital state [1] [2], despite significant studies of (and controversies on) the cycloidal orbits. The predominance of a Brillouin-like flow was subsequently demonstrated in virtually all particle-in-cell (PIC) simulations of crossed-field diodes [3] [4] [5] [6] [7]. However, the fundamental theory of Brillouin flow is always formulated in 1D. This paper reports a detailed study of 2D Brillouin flow in the presence of a step discontinuity at the anode boundary. To the authors' knowledge, no theory exists for this geometry.

The 1D Brillouin flow theory is not as straightforward as its name would suggest, even though it has been widely used to model magnetrons, MITLs and MILOs [1] [3] [6] [8] [9] [10] [11] [12]. Closed form explicit analytic solutions have only recently been systematically derived for Brillouin flows in all these geometries [11]. This theory revealed previously unsuspected properties, such as magnetic insulation for MITLs and MILOs achievable with a drive wall current less than that required at Hull-cutoff, and that MILOs typically operate very close to the Hull cutoff condition (which defines the transition between magnetic and nonmagnetic insulation), instead of the Buneman-Hartree condition (which always occurs under the condition of magnetic insulation). These new results of the 1D Brillouin flow theory are corroborated by MILO experiments that were performed at our university and elsewhere [12].

The simple 2D Brillouin flow geometry in this study consists of an abrupt discontinuity on the anode surface; the cathode surface is uniform (Fig. 1). We are motivated to study such a discontinuity in anode-cathode (AK) gap spacing, so that we may examine a situation where the diode is magnetically insulated on one side, but is not magnetically insulated on the other side of this discontinuity. Such a study, which remains to be performed, is of interest because: (1) its potential relevance to MILOs, which typically operate near the transition between magnetic insulation and conduction, i.e., near the Hull cutoff condition, as mentioned above, and (2) a sequence of such wall discontinuities on the anode, forming a slow wave structure or a choke, often appears in magnetrons or MILOs. In addition, one may argue that a steady state 2D Brillouin flow solution does not exist for the simple geometry shown in Fig. 1, even if the discontinuity in the AK gap spacing is very small, as long as it is nonzero (see Appendix A for proof of this statement). This perceived lack of a steady state solution forced us to study 2D Brillouin flow via particle-in-cell (PIC) simulation. The results, to be reported in this paper, revealed a novel bunching mechanism produced by the step discontinuity in the anode geometry.

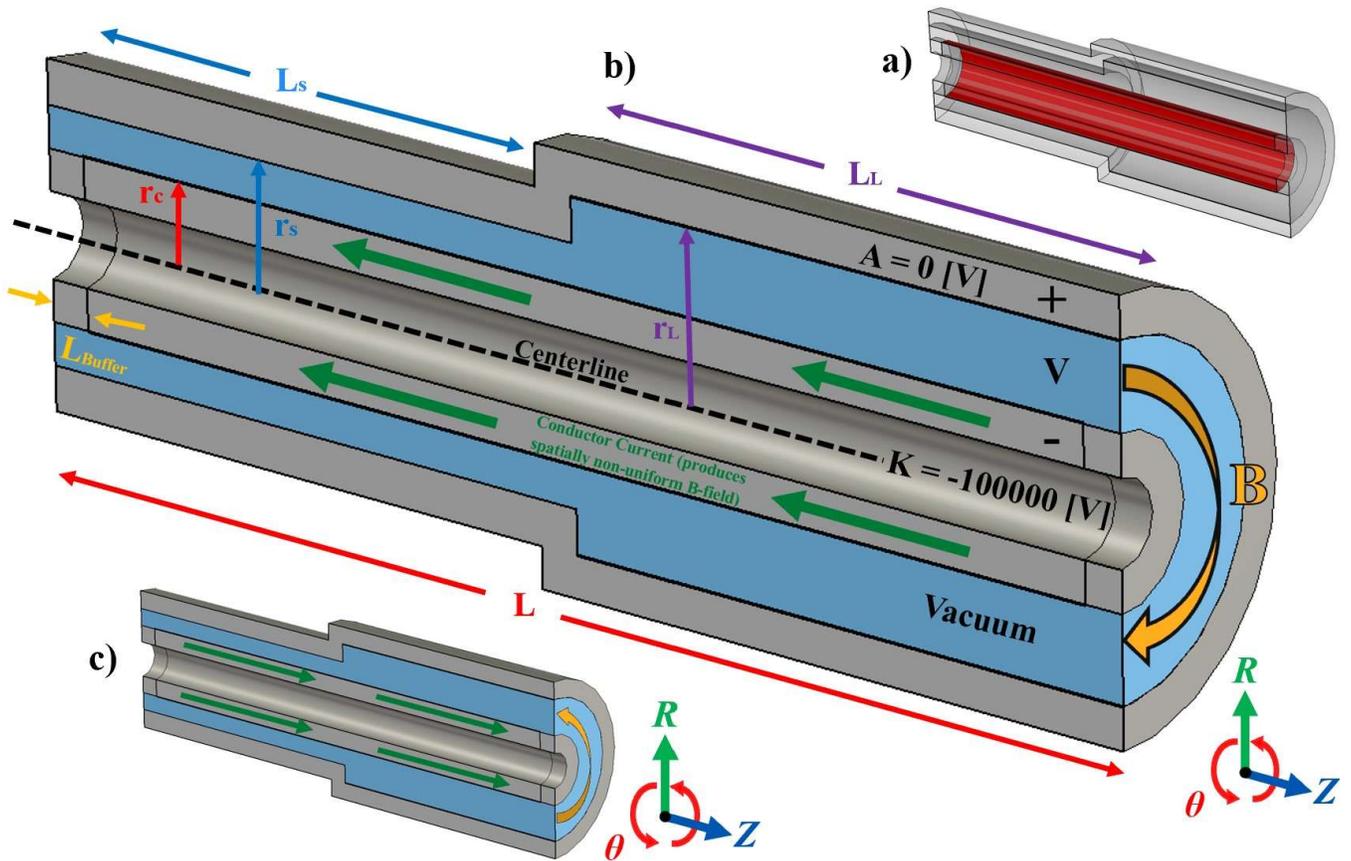

Figure 1 - Electrostatic cylindrical step discontinuity PIC model implemented in CST-PS: a) Explosive emission surface, b) nominal Brillouin electron flow from small to large gap, and c) reversed Brillouin electron flow from large to small gap. The green arrows represent the applied conductor current on the cathode that was implemented to generate the azimuthal magnetic field in the vacuum gap. The Brillouin flow is in the opposite direction of these green arrows.

Extending a classical theory from 1D to 2D is not trivial. Relevant to the present study is the 1D Child-Langmuir law (CLL [13][14]). The first notable attempt on 2D CLL was done by Luginsland et al. [15] and by Umstattd and Luginsland [16], who used PIC simulations to study the simple 2D geometry of a finite-width emission stripe on a planar diode. This initial effort, decades later, eventually led to a much deeper understanding of the underlying physics of the notoriously difficult, but practically important, problem of cathode performance (see [17] [18] for a recent account of this problem). Zhu et al. [19] most recently also used the same finite emission-stripe model to study the limiting current in a crossed-field diode with flat cathode and anode surfaces, in their extension to 2D of the earlier 1D crossed-field flow theories by Christenson et al. [4]. We remark that this 2D model of Zhu et al. [19] cannot account for the novel bunching mechanism that is reported in this paper.

We shall use the PIC solver of CST Particle Studio on a 3D, coaxial, cylindrical crossed-field diode; this ensures that the boundary conditions on the sides of the model are connected to one another, reduces unknowns, and ensures stability of the simulation and numerical results produced [20]. However, the results described in this paper are effects attributed to 2D Brillouin flow physics and are not caused by 3D effects. From the simulation results and their physical interpretation, we infer that the enhanced bunching carries the signature of the nonlinear beam modes that occur in the beam-circuit interaction in a crossed-field diode, even though the bunching mechanism is quite different from a non-magnetized diode.

This paper is organized as follows. Section II presents the simulation model. Section III presents the simulation results, in particular the circumstances under which the enhanced bunching of electrons would or would not occur. Section IV gives a qualitative description of the complex physics of this enhanced bunching. Section V concludes, including some general observations and implications. Appendix A presents a proof of the non-existence of an equilibrium Brillouin flow solution in the presence of a step discontinuity in the anode geometry (and of a smooth change in the anode geometry). The numerical issues are discussed in Appendix B.

## II. Simulation Model

In the simulation study, we first use two, separate electrostatic PIC models to set up the ideal cylindrical crossed-field device. Both models are geometrically the same (Fig. 1a); the only difference between the two models was the direction of the conductor current that generates the insulating magnetic field. In the first simulation model, the electrons flowed from the small to the large gap, generating a *clockwise* azimuthal magnetic field in the vacuum gap with respect to the z-axis (Fig. 1b). In the second simulation model (Fig. 1c), electrons flowed from the large to small gap generating a *counter-clockwise* azimuthal magnetic field with respect to the z-axis. Throughout the rest of this paper, the first case will be referred to as nominal flow and the second case will be referred to as the reversed flow. The electron explosive emission surface for both simulation models is depicted in Fig. 1a via the red coloring; note that a non-emitting buffer conductor was added on each end of the emitting surface to prevent boundary condition errors. The buffer ($L_{Buffer}$) on each end was 10 mm which reduced the overall cylindrical emission surface by 20 mm.

The dimensions of the cylindrical model are as follows: the overall length ($L$) was 300 mm, the small gap length ($L_S$) was 150 mm, the large gap length ($L_L$) was 150 mm, the radius of the cathode ($r_c$) was 20 mm, the radius to the inside edge of the small gap anode ($r_s$) was 30 mm, and the radius to the inside edge of the large gap anode ($r_L$) was 40 mm. The transition between the cylindrical, small gap and large gap regions formed an instantaneous geometric edge discontinuity at z = 0. Anode and cathode materials were selected to be perfect electrical conductors (PEC, shown in gray). The vacuum has a relative permeability = 1 (shown in light blue). Voltage was continuously applied to the anode-cathode (AK) gap throughout the entirety of the simulation runtime of 200 ns, with -100 kV applied to the cathode and the anode held at ground potential. Explosive emission parameters were set to an emission of uniform distribution with initial kinetic energy of 0.01 eV (0% velocity spread with electrons emitted perpendicular to the emission surface with low energy), an emission threshold field of 100 V/m, an emission risetime of 10 ns (slowly ramps emission to maximum over set risetime), and 115,182 emission points on the cathode. The meshing was hexahedral with equal-sized cells in all three (Cartesian) axes, and 2 million total cells. The boundary conditions on each of the ends were perfect magnetic conductors, which made the tangential components of magnetic fields, and the normal components of electric fields, zero.

Executing an electrostatic CST-PS simulation required an imported spatially non-uniform, time-independent magnetic field from the CST's Magnetostatic Solver (CST-MS). For an effective magnetic field from CST-MS, the materials had to be selected as non-PEC to solve, but a near-PEC conductivity of $10^{30}$ S/m was applied to all conductors (silver has a conductivity of ~$6.3 \times 10^7$ S/m). The meshing in the CST-MS simulation was tetrahedral with ~11.5 million mesh cells, over-resolving the largest hexahedral mesh by a factor of 5. This prevents irregular electron hotspots from propagating in the CST-PS simulation. The numerical stability and convergence of the simulation are addressed in Appendix B.

Figure 2 depicts the magnetic field generated by the CST-MS solver when a 27.2 kA conductor current was applied to the cathode (green arrow running along the cathode in Figs. 1b and 1c). This conductor current yielded the maximum magnetic field of 0.272 T on the surface of the cathode, the minimum magnetic field of 0.136 T on the large gap anode surface, and a magnetic field of 0.181 T on the surface of the small gap anode surface. This fixed cathode current of 27.2 kA (for both gaps) was chosen so that it would yield an f ~ 2 in the small gap region, and f ~ 3.6 in the large gap region, where f is the degree of magnetic insulation [12], roughly equal to the ratio of the magnetic field to the Hull cutoff magnetic field in the respective regions. Note that the use of a wall current to produce magnetic insulation mimics the MITL/MILO geometry. For the above parameters, according to the 1D theory [11], the Brillouin flow carries an electron hub current of 358 A within the small gap, and of 94 A within the large gap, both being very small compared with the current flowing on the cathode (and on the anode). In the presence of the step discontinuity in the anode wall radius, equilibrium solutions no longer exist (see Appendix A). In Section III, we report the simulation results for the nominal and reversed cases.

Since the azimuthal magnetic field is nonuniform in the cylindrical MITL/MILO geometry of Fig. 1, the Hull cutoff magnetic field is not well defined. For reference, we record the minimum current on the cathode, $I_c^{crit}$ [8] [11], that is required to provide magnetic insulation in a uniform, cylindrical MITL of cathode radius $r_c$, anode radius $r_a$, and AK gap voltage $V_a$,

$$I_c^{crit} = (8.53\ kA) \times \frac{\kappa}{\ln(r_a/r_c)} \qquad (1)$$

where $\kappa = cosh^{-1}\gamma_a = \ln[\gamma_a + (\gamma_a^2 - 1)^{1/2}]$, and $\gamma_a = 1 + V_a/511\ kV$. This Hull cutoff condition, Eq. (1), is shown [11] to be the same for the single particle orbit model (as in a vacuum diode) and for the Brillouin flow model, which assumes the space-charge-limited condition.

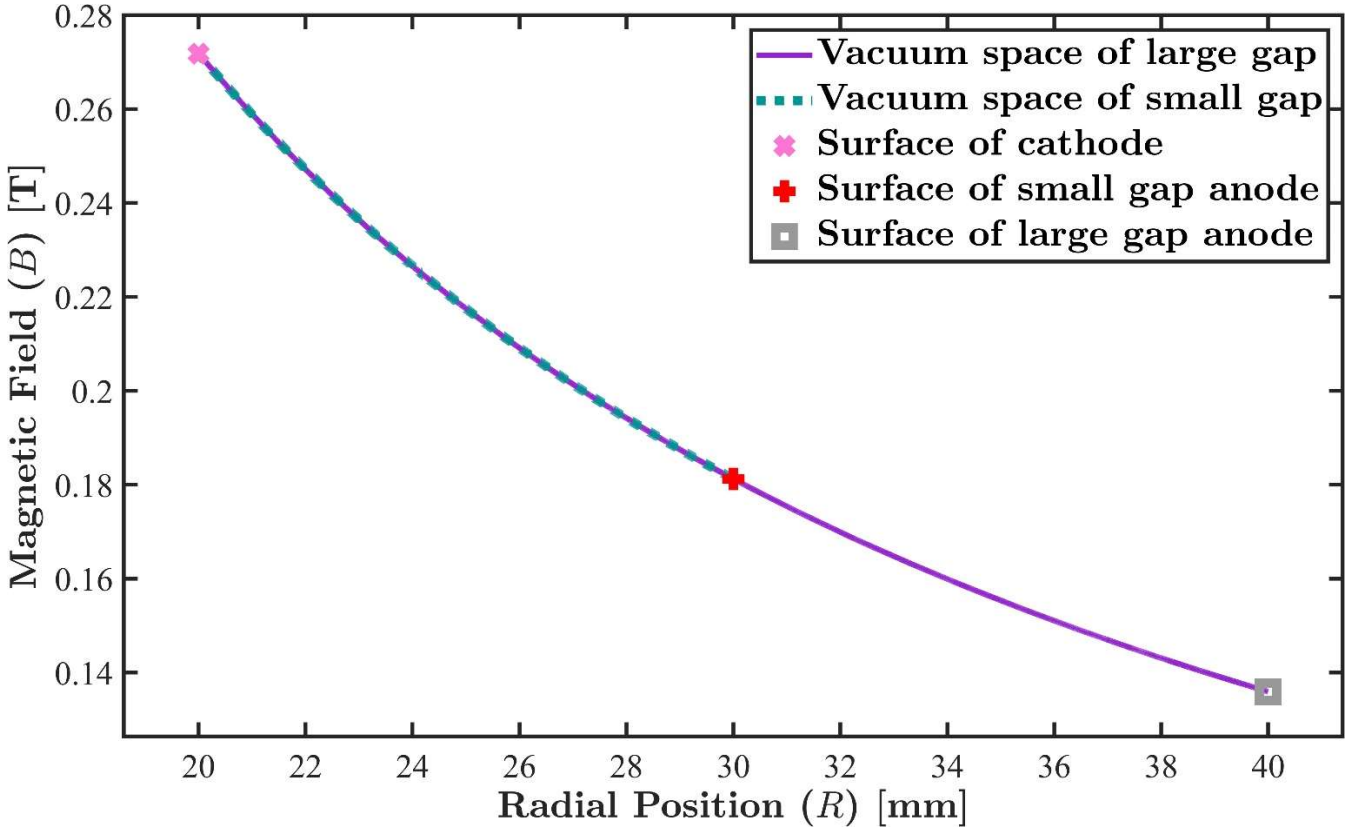

Figure 2 – Static, spatially non-uniform, azimuthal magnetic field generated by CST-MS in the cylindrical vacuum gap for a center conductor current of 27.2 kA.

## III. Simulation Results

Figure 3 shows a Brillouin-like flow profile in both gaps for the nominal case, with electron flow from the small to large gap, early (6 ns) and late (185 ns) in time. For the small gap, Fig. 3a, a clear rectilinear flow state is demonstrated with minor separation between the linear trendlines fitting to the axial velocity versus radial position of the particles for each time; this depicts normal Brillouin-like flow characteristics. However, for the large gap, Fig. 3b, the separation between the linear trendlines increases due to the significant effect of low velocity back-streaming electrons (electron axial velocity $v_z$ less than zero), providing the first indication of electron vortices forming after the wall discontinuity. Note that the data points presented in Fig. 3 are from particles on the red plane (imaged in the bottom right of each figure).

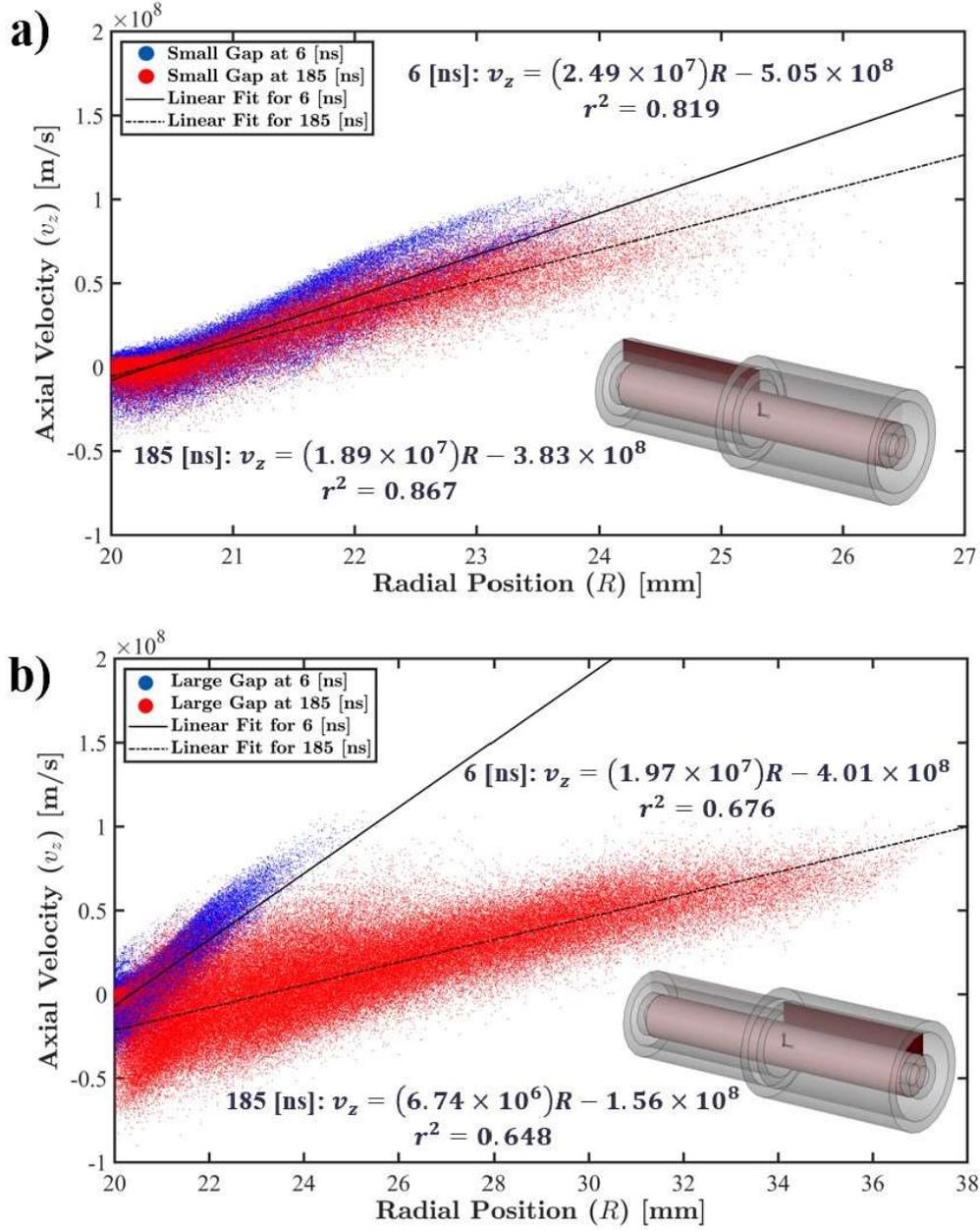

Figure 3 – Brillouin-like flow profiles for the nominal case early (6 ns) and late in time (185 ns) in the ES-PIC simulation for: **a)** the small gap on the R-z plane from z = -150 to 0 mm, and **b)** the large gap on the R-z plane from z = 0 to +150 mm. The numerical linear fits are indicated in both figures.

Figure 3b shows considerable back-streaming electrons, i.e., $v_z < 0$, especially in the large gap region. These backstreaming electrons cause the formation of vortices in the electron flow, which is evident in Fig. 4. Concomitant with the vortices is the very strong electron bunching in the electron clouds, which is also apparent in Fig. 4. Inside these intense electron clouds are virtual cathodes (see Section IV below), and these virtual cathodes in a crossed-field flow are quite different from the virtual cathodes in a non-magnetized electron flow. The complex interplay in the formation of bunching, vortices, and virtual cathodes within the crossed-field flow will be explored in Section IV. For the time being, we note that the enhanced bunching occurs only in the large gap region in the nominal case. It hardly occurs in the small gap region in the nominal case, as seen in Fig. 4 and in Fig. 5a. There is also no enhanced bunching in the reversed case, in either the large gap or the small gap region, as shown in Fig. 5b.

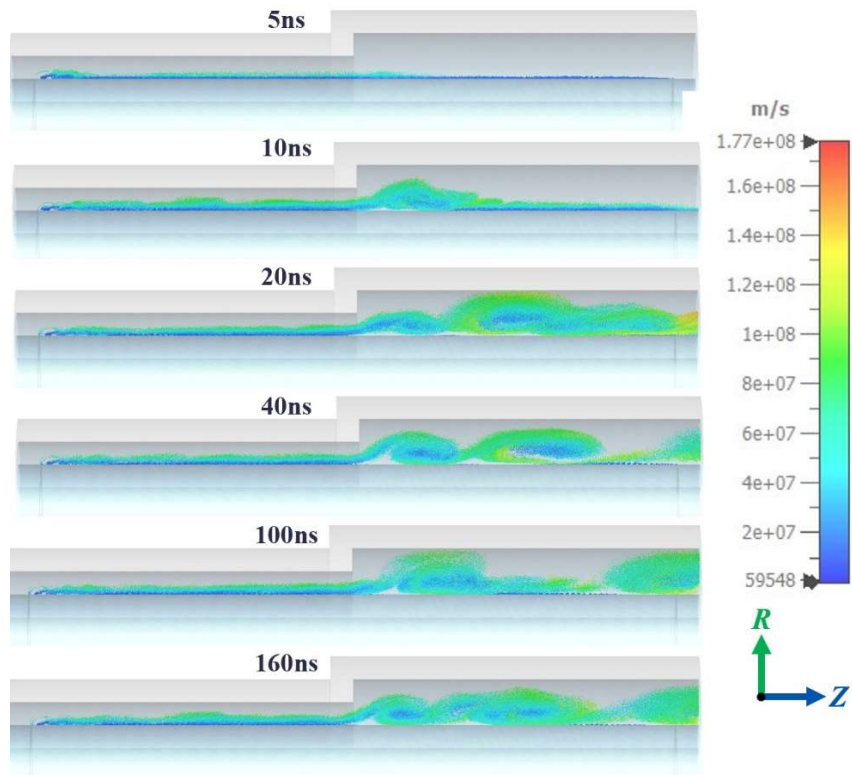

Figure 4 - 2D Cross section of the electron macroparticles' position and velocity at various times for the nominal flow case showing the evolution of the electron bunches in the large gap. The legend depicts the magnitude of the electron macroparticle velocity.

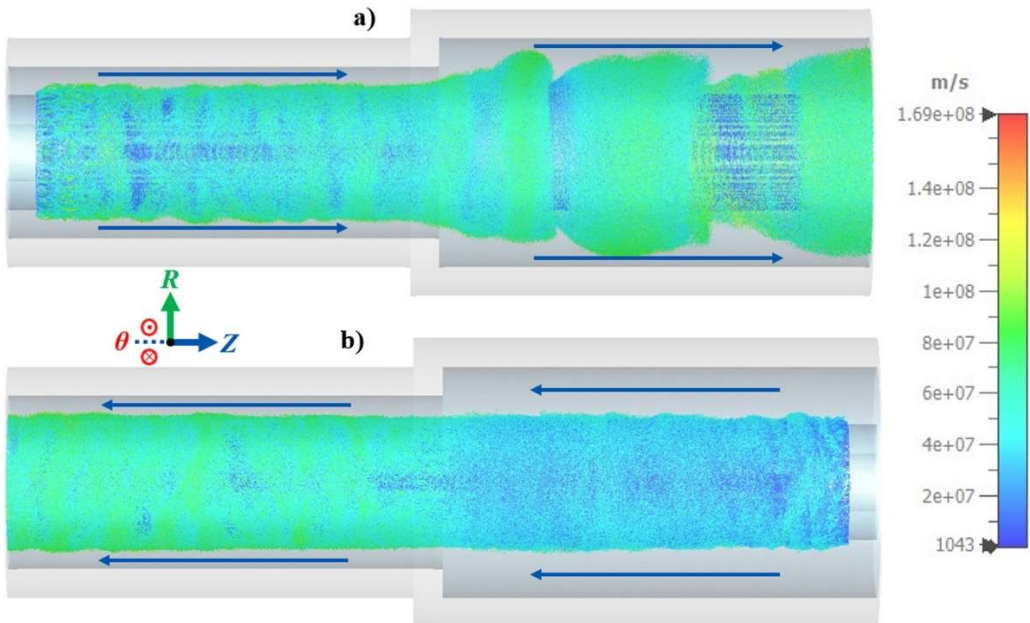

Figure 5 – Full 3D side view of the electron macroparticles' position and velocity at 190 ns for **a)** nominal Brillouin flow case, and **b)** reversed Brillouin flow case. The direction of electron flow for each case is depicted by the dark blue arrows. The legend depicts the magnitude of the electron macroparticle velocity.

The initiation of the enhanced bunching occurs very early in time (<~ 5 ns, emission risetime could play a role in the initialization timing of bunch formation) in the large gap region near the wall discontinuity ($z = 0$) for the nominal case (Fig. 4). It is explained as follows. For the nominal case, the electron (Brillouin) flow is from the small gap region to the large gap region. The magnitude of the radial electric field and, consequently, the E x B drift velocity, is higher in the small gap region. When the faster electrons from the small gap region pass the wall discontinuity at $z = 0$, they will catch up with the slower-moving electrons that are emitted from the large gap region. This leads to the pile-up of electrons slightly downstream of $z = 0$, which is faintly visible in the 5 ns frame of Fig. 4. To some degree, this electron pile-up, i.e., bunching, is similar to the all-important bunching process in a klystron in that the faster-moving electrons catch up with the slower-moving electrons. One major difference is that, in a klystron, the velocity differential is caused *temporally* by the input signal, whereas here, the velocity differential is caused *spatially* by the wall radius discontinuity. Another major difference is that the bunching in a klystron does not lead to a virtual cathode in the drift space, whereas a virtual cathode is formed here, and this will be examined more closely in Section IV. From this description, one can see why electron bunching is not initiated in the narrow gap region for the nominal case, but only in the large gap region (Fig. 5a). For the reversed case, the electron (Brillouin) flow moves from the large gap to the small gap. The slower electrons in the large gap will not catch up with the faster electrons in the small gap, explaining the absence of enhanced bunching in either region for the reversed flow (Fig. 5b).

In Fig. 5a, some helical features are seen to initiate from the electron source at the far left for the nominal case. In Fig. 5b, similar helical features are also seen to initiate from the electron source at the far right for the reversed case. A careful examination of the numerical data reveals that these helical perturbations are numerical artifacts, due to the use of hexahedral mesh. They disappear with the use of tetrahedral mesh (see Appendix B). They do not qualitatively affect the dynamics of the enhanced bunching, whose underlying physics is explored in Section IV.

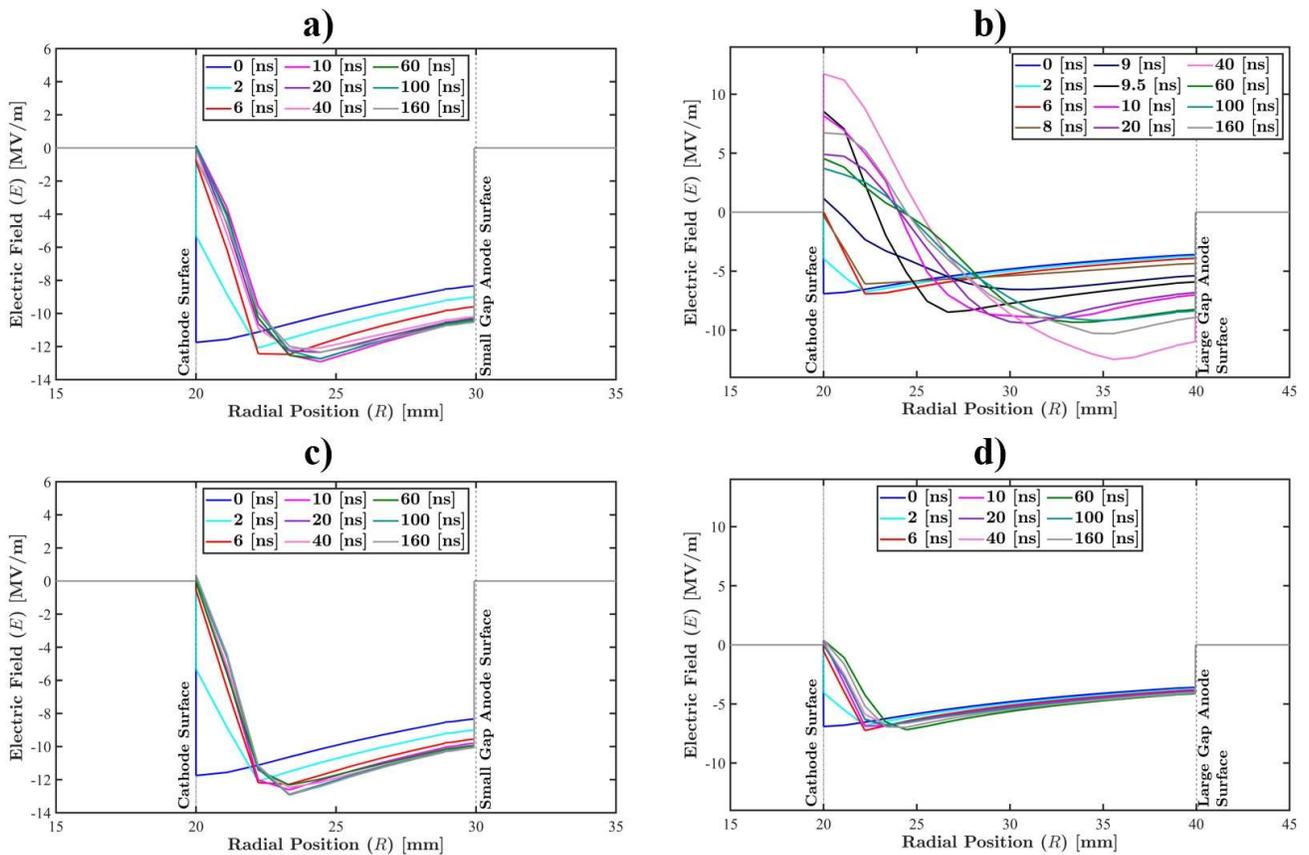

Figure 6 - Measured radial electric field at incremental time steps on a plane that is located: **a)** 30mm *before* the discontinuity (in the small gap) for *nominal* flow; **b)** 30mm *after* the discontinuity (in the large gap) for *nominal* flow; **c)** 30mm *before* the discontinuity for *reversed* flow; **d)** 30mm *after* the discontinuity for *reversed* flow.

To probe into the piling-up of electrons and the formation of the virtual cathode (or lack thereof), we show in Fig. 6 the temporal evolution of the radial electric field ($E_r$) as a function of radius for both the nominal and reversed case. Figures 6a and 6b show, respectively, the small and large gap regions for the nominal case. Figures 6c and 6d show, respectively, the small and large gap regions for the reversed case. If we assume, roughly, the presence of a virtual cathode occurs near the region where $E_r = 0$, it is seen from Figs. 6a, 6c, and 6d that virtual cathodes are formed *only* on the cathode surface, $r = r_c = 20\ mm$. The traces shown in these three figures are qualitatively similar. These virtual cathodes on the cathode surface are also qualitatively similar to the virtual cathodes that always appear on the cathode in all mildly turbulent 1D Brillouin flow simulations (see, e.g., Fig. 2c of Christenson *et al* [4]). In contrast, in Fig. 6b, corresponding to the large gap region in the nominal case, strong virtual cathodes occur at different radial positions some distance away from the cathode, during various times. This is another indication that the enhanced bunching shown in Figs. 4 and 5 is a novel phenomenon introduced by the step discontinuity in the wall radius in a crossed-field diode.

Current monitors (number of macroparticles in the vacuum gap intercepting a plane in a given timestep) are used to provide an additional diagnostic for characterizing the effect of the wall discontinuity, shown in Fig. 7 for the nominal case and in Fig. 8 for the reversed case. The most striking feature is the very large amplitude current fluctuations in the large gap region for the nominal case, as represented by the red curve in Fig. 7. The average current of 529.9 A in the large gap is significantly higher than 94 A, the 1D Brillouin flow current for the large gap [11]. In Fig. 7, the small gap current, represented by the blue curve, shows much lower-level fluctuations, with an average and fluctuating current of 480.8 ± 85 A, which is of the same order of magnitude as 358 A, the 1D Brillouin flow current for the small gap [11]. Note that the large gap current fluctuations are of a very low frequency, compared with the high frequency fluctuations seen in the small gap (the blue curve in Fig. 7). The latter high frequency fluctuations are also similar to the virtual cathode oscillations that always appear on the cathode in all 1D simulations of crossed-field diodes that end up with a mildly turbulent Brillouin flow (see, e.g., Fig. 4 of Christenson and Lau [21]).

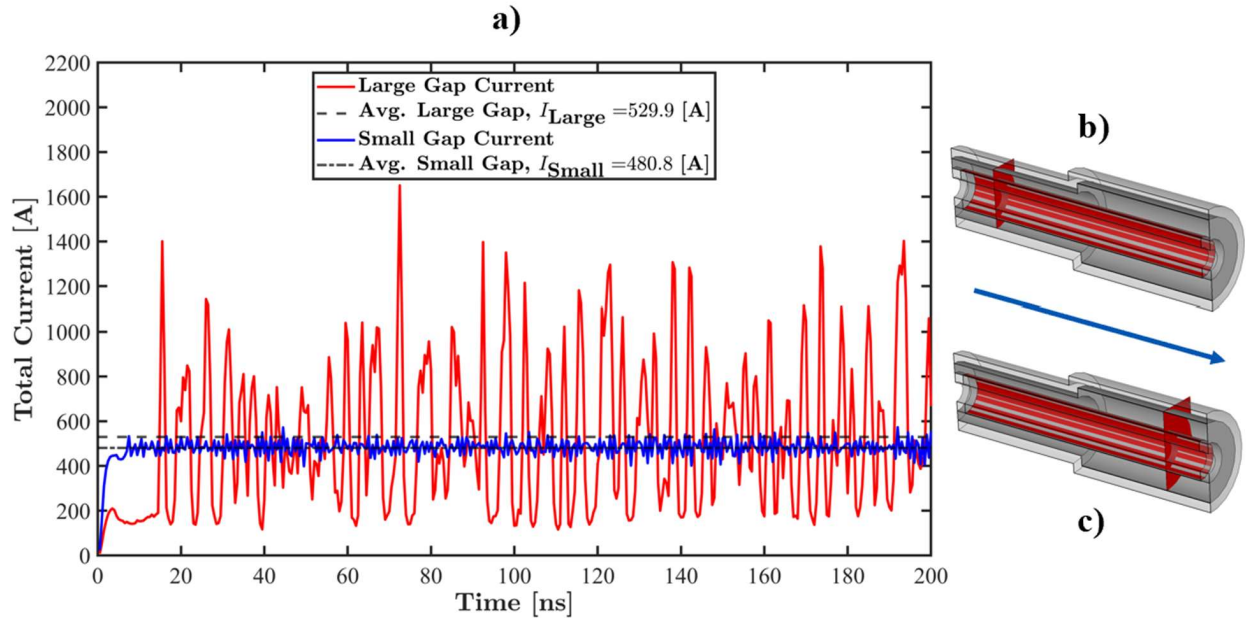

Figure 7 – Nominal flow case. **a)** Measured incident current for the electron beam in the vacuum diode gap in: **b)** the small gap at a plane placed z = -100 mm and, **c)** the large gap at z = +100 mm from the discontinuity. The location of measurement planes, and direction of electron flow, are shown in **b)** and **c)**.

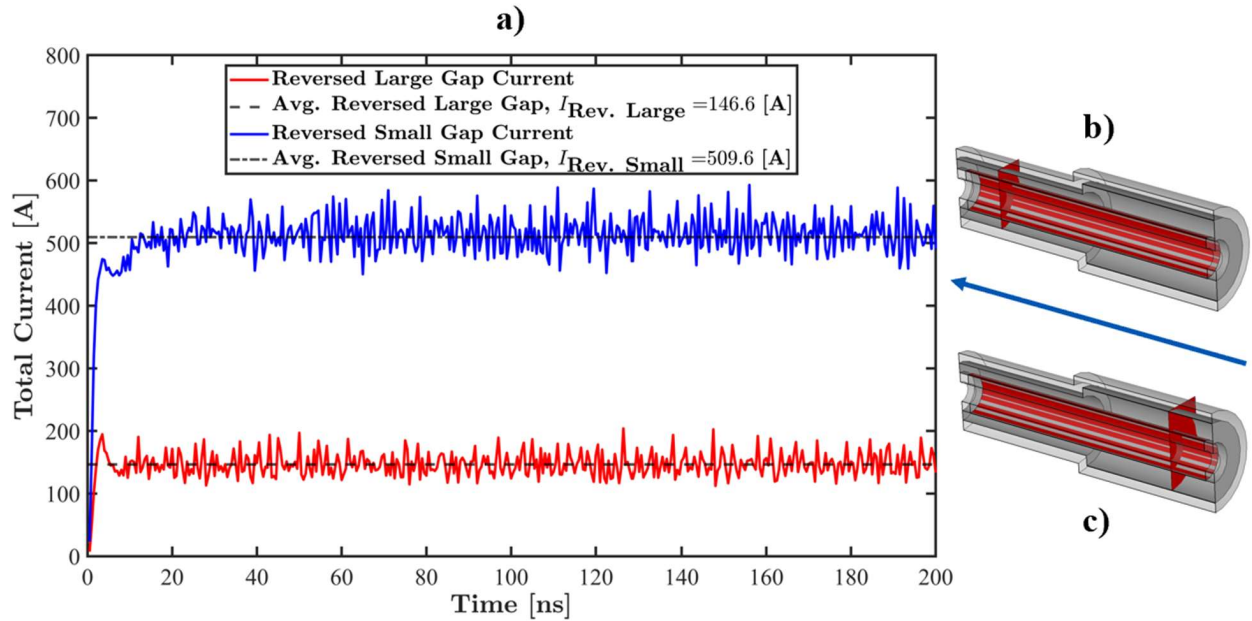

Figure 8 – Reversed flow case. **a)** Measured incident current for the electron beam in the vacuum diode gap in: **b)** the small gap at a plane placed z = -100 mm and, **c)** the large gap at z = +100 mm from the discontinuity. The location of measurement planes, and direction of electron flow, are shown in **b)** and **c)**.

For the reversed case, Fig. 8, low level fluctuations at high frequencies are seen in both the small and large gap region. In the small gap region, its average and fluctuating current of 509.6 ± 75 A is of the same order as 358 A, the hub current according to the 1D Brillouin flow theory [11]. In the large gap region, its average and fluctuating current of 146.6 ± 46 A is also of the same order as 94 A, the hub current according to the 1D Brillouin flow theory. Thus, despite the substantial wall radius discontinuity at z = 0, in the reversed case, the properties of this Brillouin flow in both regions are qualitatively similar to the 1D Brillouin flow in their respective regions. On this point, compare the similarity between Figs. 6c and 6d with Fig. 6a, and note that all these figures are qualitatively different from Fig. 6b.

The frequency spectrum of the incident large gap current for the nominal flow case is shown in Fig. 9. It shows a dominant frequency at 0.21 GHz with a large -3dB bandwidth of 0.18 GHz. These low frequency fluctuations are qualitatively different from the high frequency fluctuations, at approximately 1 GHz, that are seen in the blue curve in Fig. 7 and in both curves in Fig. 8. In comparison, the electron cyclotron frequency at the cathode surface is 7.62 GHz.

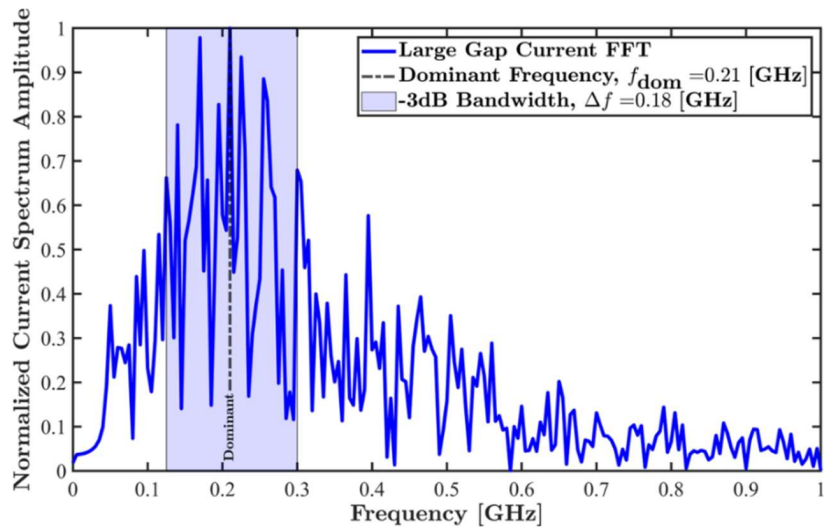

Figure 9 - Normalized fast Fourier transform, for the nominal case, of the large gap incident current spectrum for the 2 million total mesh count simulation with -3dB bandwidth.

## IV. Physics of Enhanced Bunching

The simulation results presented in Section III indicate that the enhanced bunching due to wall discontinuity results from a complex interplay of initial crowding of the electron flow, subsequent vortex formation, creation of a crossed-field virtual cathode, and the priming of a newly created bunch by its downstream neighbor. In this section, we qualitatively describe these physical processes. We focus mainly on the large gap in the nominal case where the enhanced bunching is observed.

First, the faster electron beam in the small gap will overtake the slower electron beam in the large gap near the wall radius discontinuity, resulting in the initial spatial bunching, as shown in Fig. 10a. This initial bunching quickly developed into a virtual cathode shown in Fig. 10b. The electric field pattern around this single virtual cathode is shown in Fig. 11a. Figure 11b shows the appearance of two (2) virtual cathodes that developed at a later time. The virtual cathode is defined here as the location where the total electric field is zero in the gap and is represented by the pink curve in Fig. 11a and Fig. 11b.

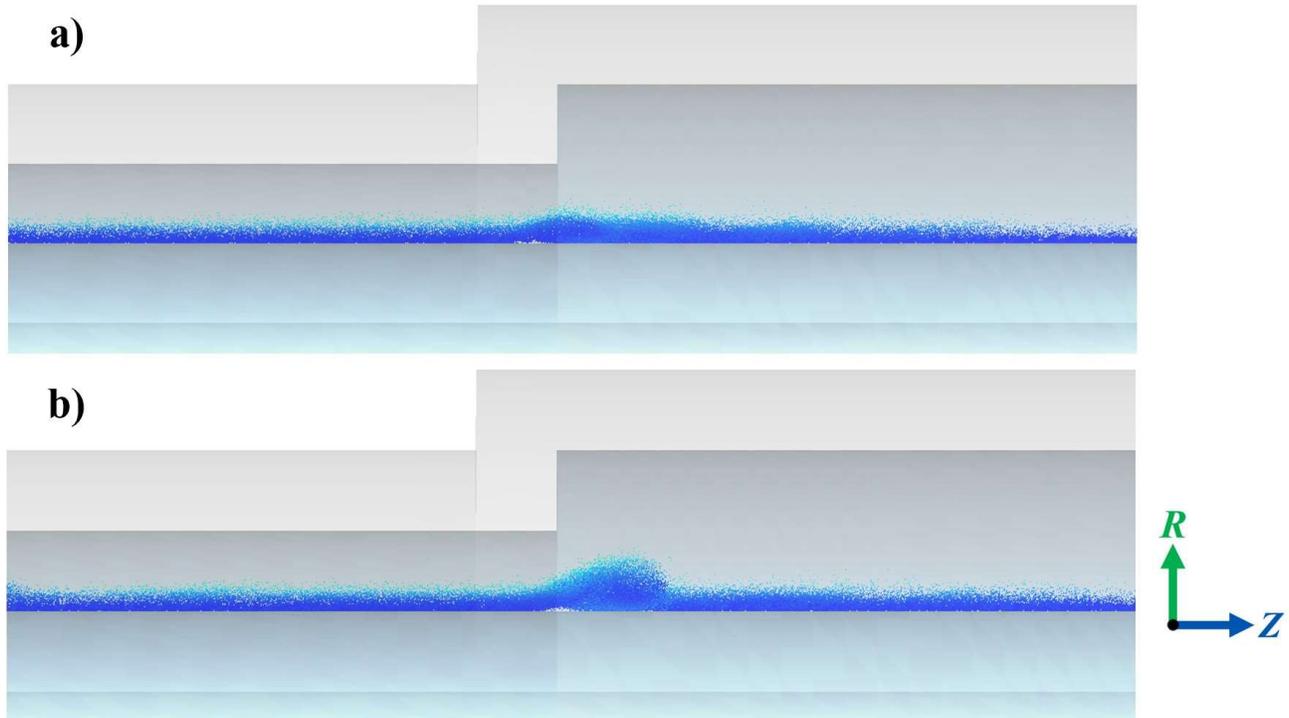

Figure 10: The electron beam is travelling in the +z direction. The E x B velocity of the electron beam in the small gap is greater than the large gap electron beam E x B velocity, which results in spatial bunching at the wall radius discontinuity seen by the 2D particle position and energy monitor at (a) 6.5 ns where the initial electron bunching can be seen at the discontinuity, and at (b) 7 ns where the initial bunch has developed, and a single large virtual cathode has formed.

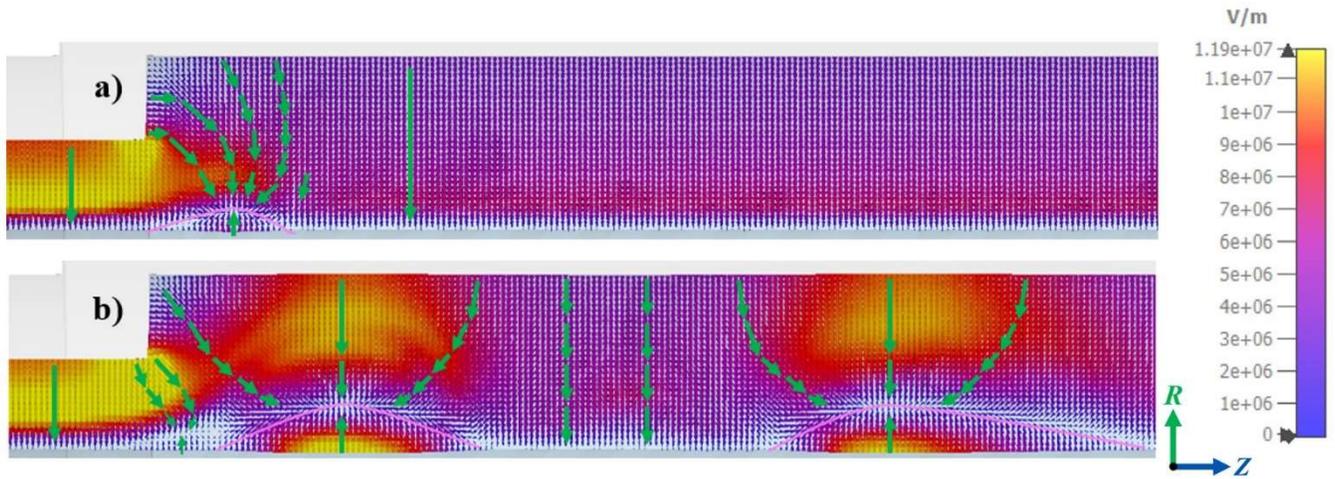

Figure 11: The electric fields in the gap near the discontinuity are depicted in green. (a) The electric field is altered by the spatial bunching where a single, large virtual cathode is formed from the initial bunch. Notably, the electric field changes direction below the virtual cathode (shown by the pink curve), thus the E x B in that region would be in the –z direction and vortical motion will occur around this virtual cathode. (b) Development of the virtual cathode when two bunches are present.

The vortex motion shown in the bunches of Fig. 12 is predicted by the Lorentz force. When the large virtual cathode forms, the local electric field shape changes considerably as seen in Fig. 11. The electric field lines point towards the virtual cathode which results in a curved electric field. From this new electric field shape, the motion of the electron beam depends on the position of the particle with respect to the electric field, and on the velocity of the particle. Assuming an initial Brillouin Flow profile, the particle trajectories around the virtual cathode will be vortical and back-streaming (electron flow in the -z direction) is possible below the virtual cathode where the electric field now points in the +R direction and the E x B direction is opposite to the equilibrium drift direction (Fig. 11).

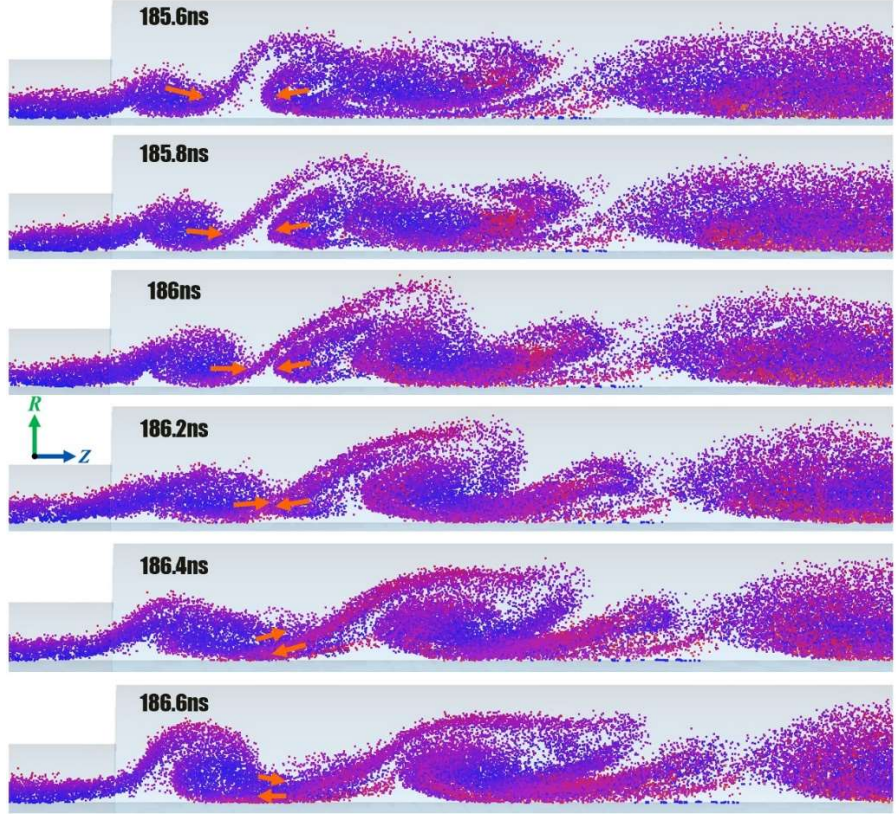

Figure 12: Particle monitors of the electron bunch formation and dynamics at the nonlinear, quasi-steady state. An electron bunch further downstream (in the +z direction) will impact the formation and dynamics of the subsequent electron bunch. The back-streaming (in the -z direction) electrons interact with the electron beam from the small gap and an increase in local space charge is observed, resulting in the formation of another electron bunch which interacts with the downstream bunch. The short orange arrows depict electron motion. This graph shows the entire axial length (150 mm) of the large gap.

As the vortical electron bunch moves axially down the gap in the +z direction, a portion of the electron bunch is back-streaming (moving in the –z direction) and will eventually interact with the electron beam upstream (to the left of back-streaming electrons) that is moving in the +z direction, which is seen in Fig. 12. The electron beam upstream of the electron bunch will begin to overtake the back-streaming electrons from the previous bunch and the local space charge will increase where the electron beam and the back-streaming electrons meet. This leads to the formation of another large virtual cathode that is shown in Figure 11b. Another vortical electron bunch will form due to the existence of the large virtual cathode in the manner described earlier. Therefore, each bunch helps prime the subsequent bunch and the back-streaming electrons from each bunch impact the formation and dynamics of the next bunch. The large virtual cathode formed by the electron bunch also prevents electron emission and a reduction in electron density can be observed in the wake of the initial bunches (Figure 12) which results in discrete bunches.

The time-resolved dynamics depicted in Fig. 12 show that the developed, mature bunches are convected downstream with a phase speed ($\omega/k_z$) on the order of $\omega/k_z \sim 1.8 \times 10^7 m/s$. In an elementary theory in plasma physics, this phase speed is also the drifting velocity of the beam in the description of the beam mode,

$$\omega \approx k_z v_z. \qquad (2)$$

Near the center of the bunches, which is located about 8 mm radially into the large gap (i.e., at $r \sim 28\ mm$) according to Fig. 12, the electron speed ($v_z$) is indeed of order $1.8 \times 10^7 m/s$ at late time (185 ns), according to Fig. 3b. Since Fig. 12 gives an axial bunching wavelength $\lambda_B \cong \lambda_z = 2\pi/k_z \sim 65\ mm$, this gives $k_z \sim 97 m^{-1}$. Equation (2) then gives the beam mode frequency ($\omega/2\pi$) on the order of 0.28 GHz. Comparing this number with the spectrum shown in Fig. 9, we see that the large-amplitude, low frequency current oscillations observed in the large gap in the nominal case carry the signature of the nonlinearly developed beam mode in the beam-circuit interaction that gives rise to the enhanced bunching and the crossed-field virtual cathode.

Because of the predominant importance of E x B drift in *all* physical processes described in this section, we see that the novel bunching mechanism described here, together with the associated formation of huge virtual cathodes within the gap, are qualitatively different from a non-magnetized diode. These virtual cathodes are also different from the ubiquitous, high frequency, virtual cathodes that always appear on the cathode surface of a crossed-field diode [21]. The virtual cathode in the 1D simulations by Christenson and Lau [21] was much milder because their 1D geometry excluded the piling up of electrons due to a change in the AK gap spacing, a process critically examined in this section. Furthermore, the bunching mechanism described here, unlike in crossed-field microwave devices, does not require a slow wave structure. The common interpretation of the spokes in a magnetron or MILO is that they arise from the synchronous interaction, as well as a self-focusing effect. Both are caused by the slow wave structure. Since the simple geometry presented here does not have a slow wave structure, the spoke-forming mechanisms are absent in our paper. Thus, this enhanced bunching mechanism is different from the resonant, spoke-forming mechanisms in magnetrons and MILOs.

## V. Concluding Remarks

In this paper, we analyze the evolution of Brillouin flow when there is a simple step change in the anode wall radius. There is no steady state Brillouin flow solution in this case even if the electron flow on both sides of this discontinuity is significantly magnetically insulated. We show that enhanced bunching occurs when the electrons flow from the region of high E x B drift velocity to the region of low E x B drift velocity, and that it occurs only in the low E x B drift region. The bunch initiation comes from the piling up of electrons due to the velocity differential on different parts of the electron flow, somewhat analogous to the ballistic bunching in a klystron. This ballistic bunching has been described in terms of the "beam mode" in the klystron literature. After this bunching initiation, the subsequent development of electron bunching differs substantially for the crossed-field diode. Vortex motion and virtual cathodes are formed due to the E x B drift, which is absent in a klystron. We probed into the complex processes and found that the fully developed bunches carry the signature of the above-mentioned beam mode, albeit highly developed nonlinearly. There is no enhanced bunching in the high E x B drift region, nor in all regions if the electron flow is from the low E x B drift region to the high E x B drift region. In the latter cases, the electron flow resembles that in a uniform, 1D crossed-field diode, namely, an ideal 1D Brillouin flow in a mildly turbulent background. The high frequency fluctuations also appear similar.

The underlying physics explored in Section IV suggests that strong vortices, and the virtual cathodes embedded therein, would occur in general if the Brillouin flow is in the direction from the high E x B drift region to the low E x B drift region. The radial MITL has this property as the electrons in the Brillouin hub flow radially inward. (See, e.g., Fig. 14 of [11] on this radial velocity differential.) Thus, strong vortices similar to those exhibited in Fig. 12 are anticipated in a radial MITL. Indeed, strong electron bunching and vortices were evident in the simulation studies by Evstatiev and Hess [7] on the MITL of the Sandia Z-machine. However, these authors' work concerned mainly the current loss via conduction current across the gap and the improvement of such simulations via different computational methodologies (decreasing the simulation time) rather than the plasma physics behind this bunching and vortex phenomenon of the Brillouin flow in the radial MITL. Here, we present plausible physical arguments for the appearance of these strong vortices, each of which might enclose a virtual cathode similar to Figs. 11 and 12. Should this be the case, we may further infer that such strong bunching and vortices would not be expected if the radial MITL operates in the "reverse polarity" where the electrons flow radially outward, from the region of low E x B drift velocity to the region of high E x B drift velocity [cf. Fig. 5b]. Finally, this paper shows that vortex formation in a crossed-field diode may be caused by the velocity differential *along* the streamlines, *possibly far more than* by the velocity differential *across* the streamlines. The latter is the well-known velocity shear that drives the Kelvin-Helmholtz instability and the diocotron instability; both of which are also known to generate vortex motions in a crossed-field flow.

The transition from magnetic insulation to nonmagnetic insulation, identified in [11] and [12] as an important issue for MILO operation, remains to be studied.

## VI. Acknowledgements

We would like to thank Matthew Hopkins, and Marco Acciarri for help in understanding the limitations of PIC, and Dion Li for his earlier participations in the simulation of 2D Brillouin flow in Cartesian geometry. We would also like to thank John Luginsland for stimulating discussions on the Brillouin flow PIC simulations. This work was supported in part by the Air Force Office of Scientific Research (AFOSR) Grant No. FA9550-20-1-0409 and No. FA9550-21-1-0184, and in part by the Office of Naval Research (ONR) Award No. N00014-23-1-2143. We acknowledge the computational resources and services provided by Advanced Research Computing (ARC) of Information and Technology Services (ITS) at the University of Michigan, Ann Arbor.

## VII. DATA AVAILABILITY

The data that support the findings of this study are available from the corresponding author upon reasonable request.

## APPENDIX A. PROOF OF NON-EXISTENCE OF EQUILIBRIUM BRILLOUIN FLOW SOLUTION IN THE PRESENCE OF A STEP DISCONTINUITY IN THE ANODE GEOMETRY

For simplicity, consider a nonrelativistic, planar crossed-field diode with a step discontinuity on the anode geometry at z = 0. For z < 0, the gap spacing is $D_1$, and for z > 0 the gap spacing is $D_2$ ($D_1 \neq D_2$). For all z, the gap voltage is a constant V and the magnetic field is another constant B which is assumed to be higher than the Hull cutoff magnetic field in either region. For this geometry, we furnish the nonexistence proof by first assuming that an equilibrium Brillouin flow solution exists, and then show that a contradiction results.

If an equilibrium Brillouin flow solution exists, all flow quantities are time independent and the continuity equation implies,

$$\frac{\partial \rho}{\partial t} = -\nabla \cdot \boldsymbol{J} = 0, \quad (A1)$$

where $\rho$ is the charge density and $\boldsymbol{J}$ is the current density of the Brillouin flow electrons. At a distance very far from the discontinuity (z = 0), the effect of the discontinuity vanishes, and the equilibrium Brillouin flow must carry a total electron current, $I_1$ and $I_2$, in the two regions, respectively with gap spacings, $D_1$ and $D_2$. Since $D_1 \neq D_2$, we have $I_1 \neq I_2$ as V and B are the same in both regions. This means that $\nabla \cdot \boldsymbol{J} \neq 0$, which contradicts Eq. (A1). Thus, a step discontinuity in the boundary, regardless how small, as long as it is nonzero, prevents the existence of equilibrium flow solution, strictly speaking.

Using the above proof, one may similarly argue that if there is a change in the gap spacing, even if this change is *gradual*, a time-independent Brillouin flow solution does not exist.

## APPENDIX B. ASSESSMENT OF NUMERICAL STABILITY

The veracity of simulated plasma phenomena has been a topic of debate since the inception of PIC in the early 1970s, major contributors to the field being Buneman, Boris, and Birdsall [22-24]. Many numerical errors can arise during a simulation that form artificial instabilities when operating outside or near the limitations of PIC simulations [25]. These artificial instabilities are always present; however, the growth rate of such instabilities can be decreased to the point where it can be ignored by operating within the PIC limitations. The simulations discussed in this paper fall into the collisionless, non-neutral plasma regime since the number of particles in the Debye sphere was large, and the length of the model was also large compared to the Debye length. The only particles injected into the AK gap were electrons, simplifying possible numerical instabilities [26]. For collisionless, non-neutral plasmas, only four simulation parameters are of interest in determining the viability of such simulations: the spatial resolution ($S_r$), the temporal resolution ($T_r$), the number of macroparticles per cell (*NMP*), and the macroparticle speeding criteria (*MPV*) which is synonymous with the Courant-Friedrichs-Lewy (CFL) Condition [26]. Spatial resolution describes how resolved the Debye length ($\lambda_D$) is with respect to the mesh cell size ($\Delta x$), $S_r = \lambda_D/\Delta x$. Temporal resolution describes how resolved the electron plasma frequency ($f_{pe} = 1/t_{pe}$) is when compared to the simulation time step ($\Delta t$), $T_r = t_{pe}/\Delta t$. Macroparticle speeding describes how resolved each particle is within each cell for a given time step ($v_{max}$ = max electron velocity), $MPV = v_{max}\Delta t/\Delta x$. The *MPV* must always be less than 1 for any simulation; if *MPV* is not less than 1, then the particles will transit a mesh cell before the next time step, invalidating the simulation. The number of macroparticles per cell provides insight into how the solver interprets charge and fields on each of the nodes between mesh cells. As macroparticles move from one cell to the next they cause a noise spike on the mesh cell boundary; more macroparticles in each cell will smooth and reduce these noise spikes [25]. To reduce the inherent numerical noise and error of PIC simulations, the spatial and temporal resolution should be greater than 1, and the macroparticle speeding should be less than 1. Other simulation parameters might have to be considered depending on the number of plasma/gas species, the plasma temperature, model size, and strength of the magnetic field, as described by Parker, Birdsall, Horký, and Ueda [27-30].

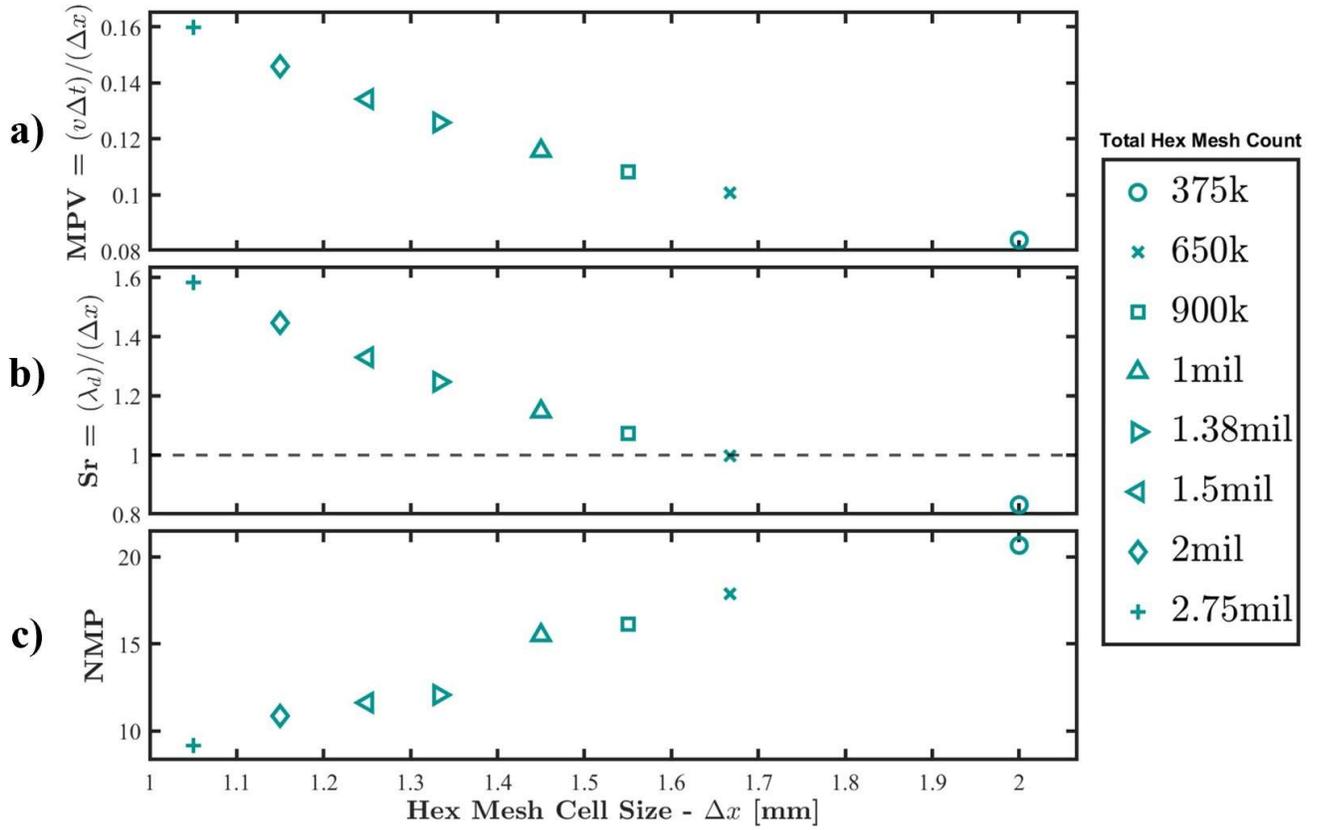

Figure 13 - ES-PIC simulation parameters: **a)** macroparticle speeding metric (*MPV*), **b)** spatial resolution ($S_r$), **c)** number of macroparticles per mesh cell of occupied vacuum volume (*NMP*).

To determine the reliability of the simulations in producing "real" plasma physics, a sweep of the total hex mesh count (or the hex mesh cell size) was conducted from 375 thousand to 2.75 million mesh cells (2 mm to 1.05 mm mesh cell size), 3 of the 4 simulation parameters of interest are presented in Fig. 13. For all the simulations present in Fig. 13, the simulation time step was kept at 2 ps for a 200 ns total simulation time producing a temporal resolution ($T_r$) of 4.46 for the applied voltage of -100 kV. Also, the *MPV* for all the simulations was below 1. At and below 650k mesh cells, the simulations were spatially unresolved. The *NMP* of the simulations ranged from 9 to 21 in the *occupied* vacuum gap volume of the simulations examined. There is still active debate on the necessity of a large *NMP* for collisionless, non-neutral (electron only) simulations [26]. As the number of mesh cells increased (mesh cell size decreased), the total simulation time, and thus computational resources, increased drastically from about 8 hours (375k cells) to 8.5 days (2.75M cells) when running on the Great Lakes High Performance Computing Cluster (GL HPC, see Acknowledgments). For 2M mesh cells (1.15 mm mesh cell size) the spatial resolution was 1.45, *MPV* was 0.146, and *NMP* was 11.6, so all 4 simulation parameters are within PIC limitations and can be considered reliable for 2M mesh cell simulation. Therefore, this paper utilized only the 2M total mesh cell simulations for the nominal and reversed simulations that were discussed in the main text.

Helical perturbations (numerical instability) in the Brillouin flow were produced as a side effect of coarse tessellation from the hexahedral mesh on the cylindrical surfaces of the model. These perturbations were eliminated using an unstructured tetrahedral mesh on the cylindrical 3D model which removed the sharp edge effects [31]. Due to computational cost of tetrahedral meshing of electrostatic PIC simulations with CST-PS at large mesh counts (>500k), the hexahedral mesh simulations were utilized over the tetrahedral mesh simulations, as shown in Fig. 5.

To analyze the frequency content, a highpass filter with a 0.05 GHz pass frequency was used to eliminate the low frequency noise on the signal. However, the FFT cannot be used effectively to find the mean local time between current peaks ($t_{mlt}$) depicted in Fig. 9 because the simulation's large-incident-gap current signal does not have large enough sample size to fully resolve the frequency spectrum of operation. Also, the signal is poorly resolved according to the Nyquist theorem because of the computational cost of the simulation. This poor resolution stems from the restriction that these simulations utilized 0.5 ns time step on the particle position monitor; hence, one can only resolve up to 1 GHz, when assuming 2 datapoints per wavelength. That is why the rudimentary approach of finding the $t_{mlt}$ via current peak prominence, a means of determining local extrema in datasets, was used over the FFT. Some frequency information may be lost with the current peak prominence, but due to under-resolved signal, the FFT could not be implemented properly to generate $t_{mlt}$ values.

Mean local time between incident peak large gap current ($t_{mlt}$) using the prominence method for various simulation hex mesh cell sizes for the nominal Brillouin flow case is presented in Fig. 14. This shows a convergence of $t_{mlt}$ to ~3.5 ns as the simulation mesh cell size decreases. This indicates that there is on average 3.5 ns between one electron bunch and the next for a converged simulation of this system.

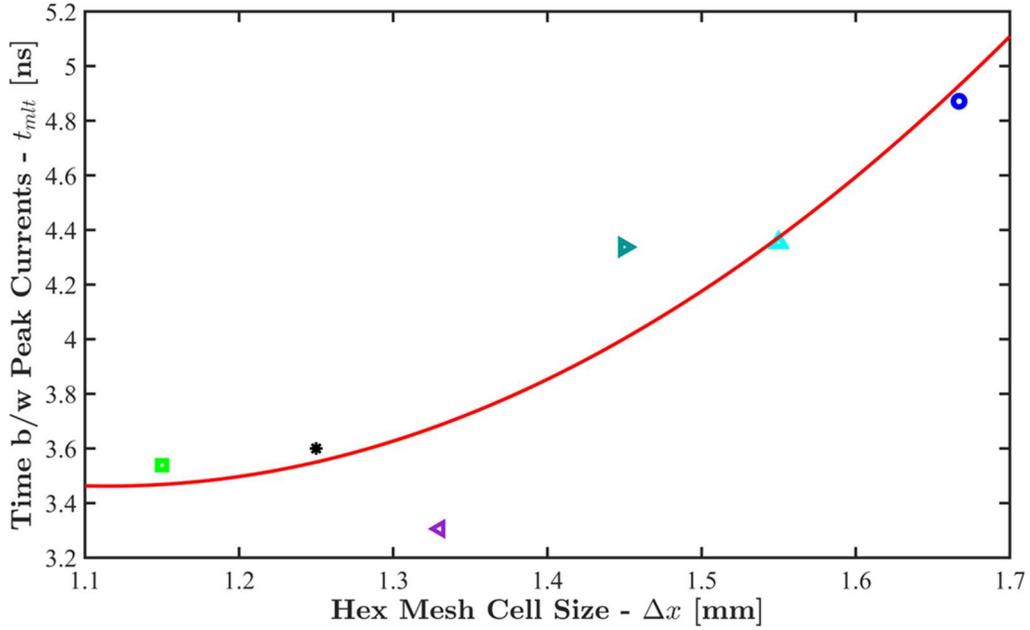

Figure 14 - Mean local time between peak currents ($t_{mlt}$) using peak current prominence method with various spatial mesh sizes. This demonstrates that the ES-PIC simulations converged at higher total mesh counts or smaller mesh cell sizes.

Finally, Fig. 15 depicts the gradual convergence of bunching wavelength ($\lambda_B \sim \lambda_z$) as a function of decreasing mesh cell size, similar in form to Fig. 14.

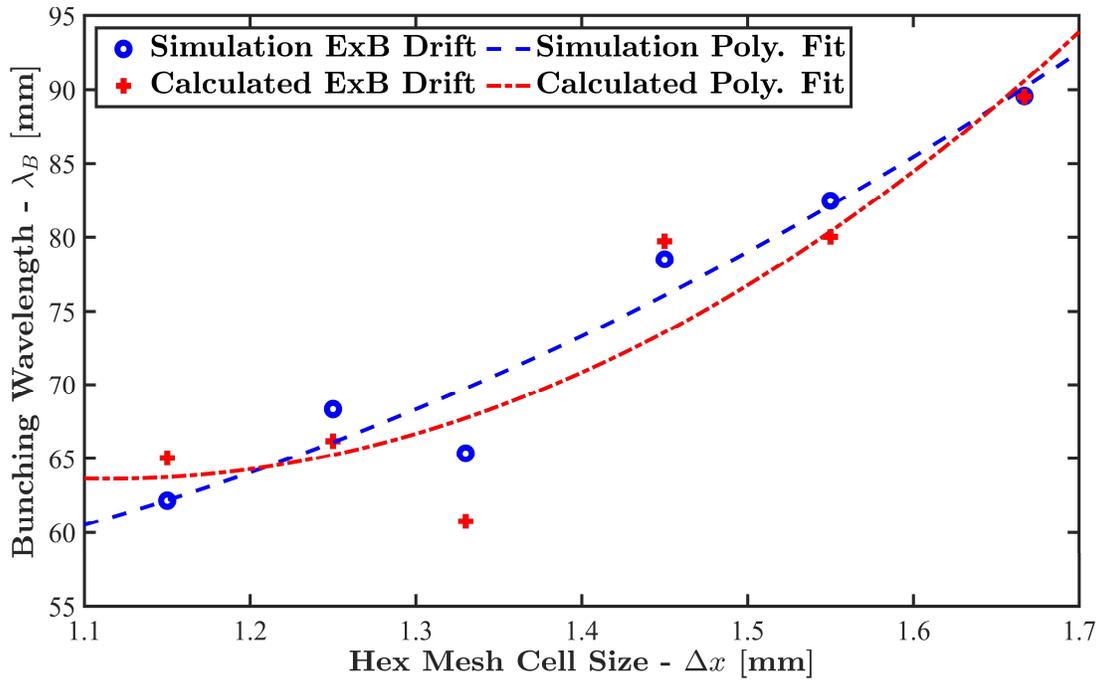

Figure 15 - Respective bunching wavelength ($\lambda_B$) for ES-PIC simulations with various spatial mesh sizes. For the simulation E x B drift datapoints, the average axial velocities of the large gap at 190 ns were used for each simulation to calculate the bunching wavelength (1.5 million mesh had to use 192 ns due to large nulls from uncommon spoking behavior at 190 ns). For the calculated E x B drift datapoints, the approximate drift was found via (electric field)/(max magnetic field on cathode surface).